\documentclass[aps,prl,twocolumn,superscriptaddress,showpacs]{revtex4-1}
\usepackage{fullpage}
\usepackage{mathrsfs} \usepackage{amssymb} \usepackage{graphicx,color} \usepackage{bbm} \usepackage{multirow}
\usepackage{bm} \usepackage{mathrsfs} \usepackage{commath} \usepackage{stmaryrd} \usepackage{color}
\usepackage[sort&compress]{natbib}
\usepackage{subfigure}

\usepackage[latin1]{inputenc}
\usepackage{ulem}
\usepackage{amsmath}
\usepackage{amssymb}
\usepackage{bm}
\usepackage{graphicx}
\usepackage{float}

\usepackage{hyperref}
\hypersetup{
  colorlinks   = true,    
  urlcolor     = blue,    
  linkcolor    = blue,    
  citecolor    = blue     
}

\newcommand{\beq}{\begin{equation}}
\newcommand{\eeq}{\end{equation}}

\begin{document}

\title{Bath induced phase transition in a Luttinger liquid}

\author{Saptarshi Majumdar}
\affiliation{Universit\'{e} Paris Saclay, CNRS,LPTMS, 91405, Orsay, France}

\author{Laura Foini}
\affiliation{IPhT, CNRS, CEA, Universit\'{e} Paris Saclay, 91191 Gif-sur-Yvette, France}

\author{Thierry Giamarchi}
\affiliation{Department of Quantum Matter Physics, University of Geneva, 24 Quai Ernest-Ansermet, CH-1211 Geneva, Switzerland}

\author{Alberto Rosso}
\affiliation{Universit\'{e} Paris Saclay, CNRS,LPTMS, 91405, Orsay, France}

\begin{abstract}
We study an XXZ  spin chain, where each spin is coupled to an independent ohmic bath of harmonic oscillators at zero temperature. Using bosonization and numerical techniques, we show  the existence of two phases separated by an Kosterlitz-Thouless (KT) transition. At low coupling with the bath, the chain remains in a Luttinger liquid phase with a reduced but finite spin stiffness,  while above a critical coupling the system is in a dissipative phase characterized by a vanishing spin stiffness. We argue that the transport properties are also inhibited: the Luttinger liquid is a perfect conductor while the dissipative phase displays finite resistivity.  Our results show that the effect of the bath can be interpreted as annealed disorder inducing signatures of localization.
\end{abstract}

\date{\today}

\maketitle

In quantum systems, transport properties are strongly affected by the presence of quenched disorder. The most spectacular effect is the localisation of part of the spectrum of the hamiltonian, which is at the origin of a finite temperature metal-insulator transition, first predicted for free fermions by P.W.Anderson \cite{AndersonLoc}. Recently,  it has been argued that finite temperature localization can also occur in presence of interactions via the so-called many body localisation (MBL)\cite{BASKO,Valentina,MBLRev1,MBLRev2,MBLRev3,MBLRev4}.  Instead, localization is detroyed by contact with an external bath via Variable Range Hopping (VRH)\cite{Mott,GiamarchiPierre,Efros_1975}.  Indeed,  transport becomes possible by the hopping of a localised electron over a distance $r$ due to the emission or absorption of phonons.  Note that this process is feasible only for acoustic phonons which are delocalized at both donor and acceptor sites.  However for systems with single degree of freedom,  such as a single spin or a single particle,  baths can induce localisation. In particular, it has been shown that subohmic baths with harmonic oscillators at zero temperature can freeze the quantum dynamics of such systems \cite{Bray&Moore,Chakravarty,Schmid,leggettchakravarty}.
\begin{figure}[h]
\centering
\includegraphics[width=0.37\textwidth]{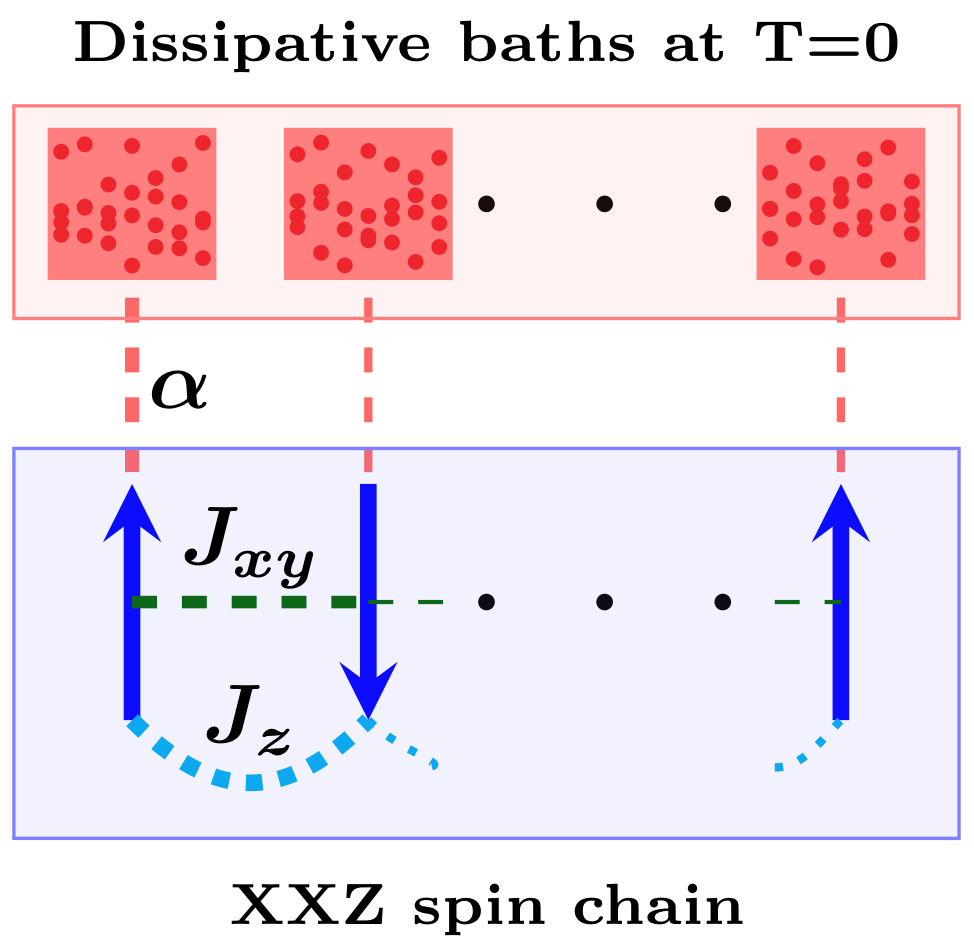}
\caption{Schematic representation of the microscopic system: An one-dimensional XXZ spin chain(blue color) with each of the spin coupled to its individual dissipative bath(red colour). The baths are described by a collection of simple harmonic oscillators kept at zero temperature. The parameter $\alpha$ is a measure of the coupling strength between the bath and the associated spin. }
\label{fig:spin_chain} 
\end{figure}
In our paper,  we  extend these works and show the existence of a genuine bath-induced phase transition in many-body systems. To avoid processes like VRH, we work with a bath that produces optical phonons.  We focus on one dimensional system to use an approach that allows us to probe very large systems and account for the effect of the bath in an exact way. In practice, we choose the system to be an XXZ spin-chain with the hamiltonian $H_S = \sum_{j=1}^N J_z S_j^z S_{j+1}^z +J_{xy} \left(S_j^xS_{j+1}^x + S_j^yS_{j+1}^y  \right)  $  and $J_z/J_{xy} \in (-1,1)$.  This model displays a gapless low energy spectrum and it is in a perfectly conducting phase known as Luttinger Liquid (LL) \cite{giamarchibook}.  Each spin $j$ of the chain is in contact with its own independent bath of harmonic oscillators with the hamiltonian $  H_{\text{B}} = \sum_{j k} \frac{P^2_{jk}}{2 m_k} + \frac{m_k \Omega_k^2}{2} X_{jk}^2 $.  A different choice for local baths was studied in \cite{cazalillalocal}. The complete hamiltonian is given by :
\begin{align}
 \begin{split}
H &= H_{\text{S}}+H_{\text{B}}+H_{\text{SB}} \\
 H_{\text{SB}} &= \sum_{j=1}^N S_j^z \sum_k \lambda_k X_{jk}
 \label{eq:spin_ham}
\end{split}
\end{align}
Note that the coupling term $h_j(t)= \sum_k \lambda_k X_{jk} $ is equivalent to a time-dependent magnetic field interacting with the spins. The time-independent limit, $h_j(t)=h_j $, corresponds to a quenched disordered magnetic field. Recent numerical simulations seem to suggest that quenched disorder can induce a finite temperature MBL transition in this model \cite{Serbyn}.  At zero temperature instead,  a localisation transition surely occurs and can be studied using Bosonization \cite{GiamarchiSchulz2} and powerful simulation techniques \cite{Laflorencie}. Here we replace the quenched disorder by an annealed disorder produced by the bath and investigate the possibility of a zero temperature localisation transition. To fully characterize the bath, we need to specify the low-frequency behaviour of the spectral function, defined as :
\begin{equation}
J(\Omega)=\frac{\pi}{2} \sum_k (\lambda_k^2/m_k \Omega_k) \delta(\Omega-\Omega_k)
\end{equation}
In general, one has $J(\Omega) = \pi \alpha \Omega^s $ for $\Omega \in (0,\Omega_D)$. Here $\alpha$ denotes the effective coupling strength with the bath, the cut-off $\Omega_D$ is the Debye frequency and $s$ sets the nature of the bath.  For our study we take $s=1$, which corresponds to an ohmic bath. A similar model was already studied by Bosonisation \cite{cazaillashort} and Monte-Carlo techniques \cite{BIBL} in a different context. However,  its phase diagram remains controversial and it is not clear how many phases appear varying the parameter $\alpha$.  Here, we introduce a novel approach, which simulates directly the bosonised action and allows us to reach large system sizes. Our results show a simple scenario of two phases with a KT transition between them.  Increasing $\alpha$, a dissipative phase with suppressed transport takes over the LL phase.

\textit{Bosonised action}: We map the chain with periodic condition into a 1D fermionic system using Jordan-Wigner transformation.  For $J_z=0$, we recover the free-fermion problem that can be diagonised in the momentum space $q=2\pi l/(Na)$ with $a$ as lattice spacing and $l \in (-N/2,N/2)$. The Fermi momentum depends on the total magnetisation $M$ of the spin chain, namely $q_F = \pi(N-M)/(2Na) $. Two cases should be distinguished : In the zero sector of magnetisation, $q_F$ is commensurate with the lattice space, while it is incommensurate for the non-zero magnetisation sector. Here we focus on the incommensurate case.  Bosonisation allows us to include both the interaction and the bath by linearizing the spectrum around $q_F$, which corresponds to the low-energy physics of the system.  Using the standard techniques (\cite{Note1}, Sec. 1), we can map the spin-chain problem into the following field theory action $S= S_{\text{LL}}+S_{\text{int}}$, where:
\begin{align*}
S_{\text{LL}} & =\frac{1}{2\pi K} \int dx d\tau \left[ \frac{1}{u} \left(\partial_{\tau} \phi(x,\tau) \right)^2 + u\left(\partial_x \phi(x,\tau) \right)^2  \right] \tag{3} 
\label{eq:S_LL}\\
S_{\text{int}}  &=- \frac{\alpha}{4\pi^2} \int dx d\tau d\tau' \frac{\cos \left(2\left(\phi(x,\tau)-\phi(x,\tau') \right)\right)}{|\tau-\tau'|^2} \tag{4}
\label{eq:S_int}
\end{align*}
Here $\phi(x,\tau)$ is a two-dimensional field living in the physical space $x \in (0,L)$ and in imaginary time $\tau \in (0,\beta)$, where $\beta$ is the inverse temperature.  Thermodynamic quantities of the spin chain can be expressed in terms of the correlation functions of the field $\phi$. In particular, the propagator $G(q,\omega_n)=\langle \phi(q,\omega_n)\phi(-q,-\omega_n)\rangle $ can be related to the susceptibility $\chi$ and spin stiffness $\rho_s$ by the two equations:
\begin{align*}
\chi &= \lim_{q \to 0} \lim_{\omega_n \to 0} \frac{q^2}{\pi^2}G(q,\omega_n)\tag{5}
\label{eq:chi}\\ 
\rho_s &=\lim_{\omega_n \to 0}\lim_{q \to 0} \frac{\omega_n^2}{\pi^2} G(q,\omega_n) \tag{6}
\label{eq:rho}
\end{align*}
Here $\omega_n=2\pi n/\beta$, $n \in (-\beta/2,\beta/2)$ are the Matsubara frequencies.  When $\alpha=0$, we recover the LL action which corresponds to an isolated XXZ spin chain and the parameters $u$ and $K$ can be directly related to $J_z$ and $J_{xy}$ either by Bethe Ansatz or by bosonisation (\footnote{See SI for additional information}, Sec. 1). In this phase, $G_{\text{LL}}(q,\omega_n) = \pi K/(\omega_n^2/u+u q^2)$, and hence $\chi= K /(u\pi)$ and $\rho_s =  u K/ \pi $. 

\begin{figure*}[ht]
\centering
\includegraphics[width=0.33\linewidth]{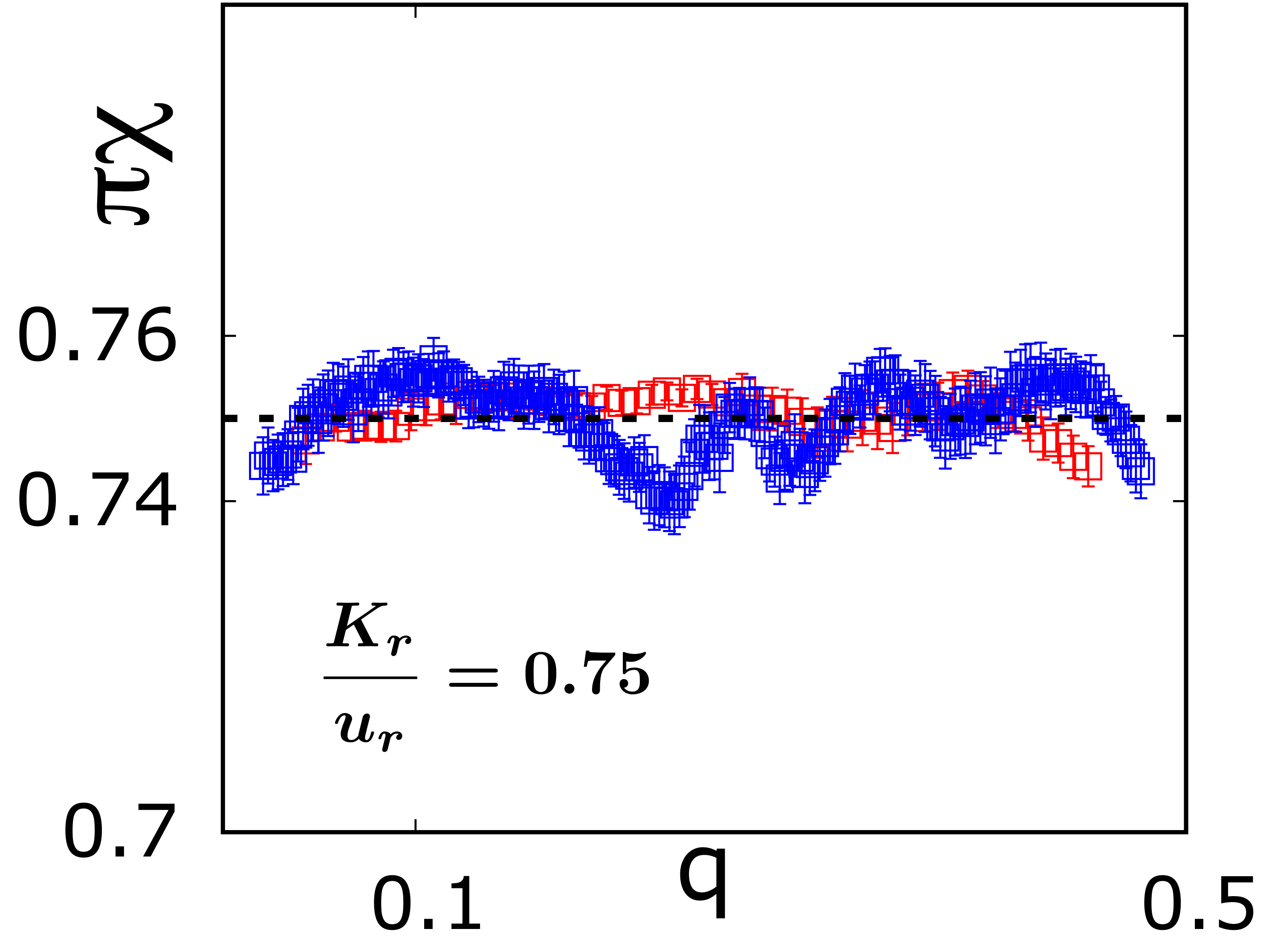}
\includegraphics[width=0.32\linewidth]{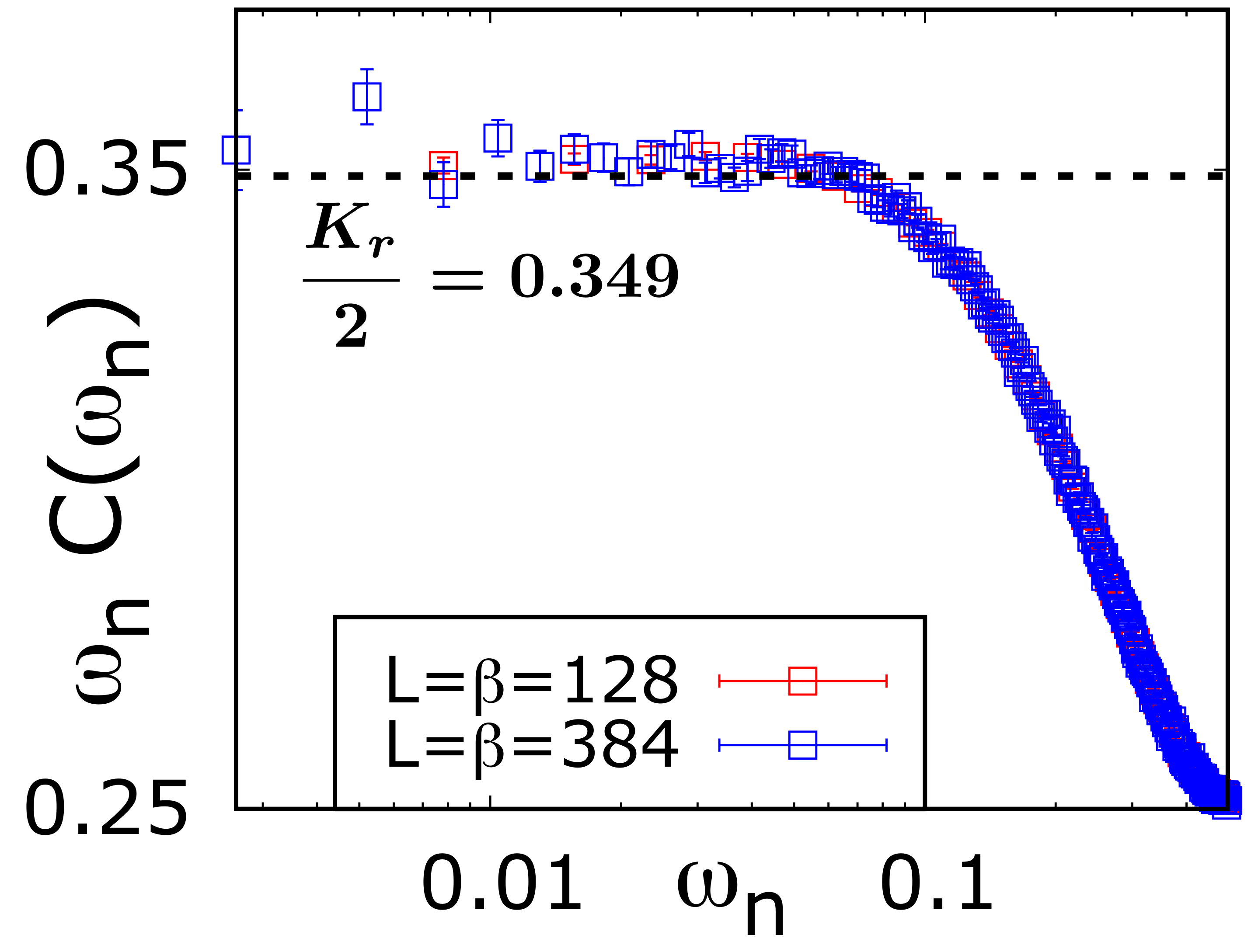}
\includegraphics[width=0.31\linewidth]{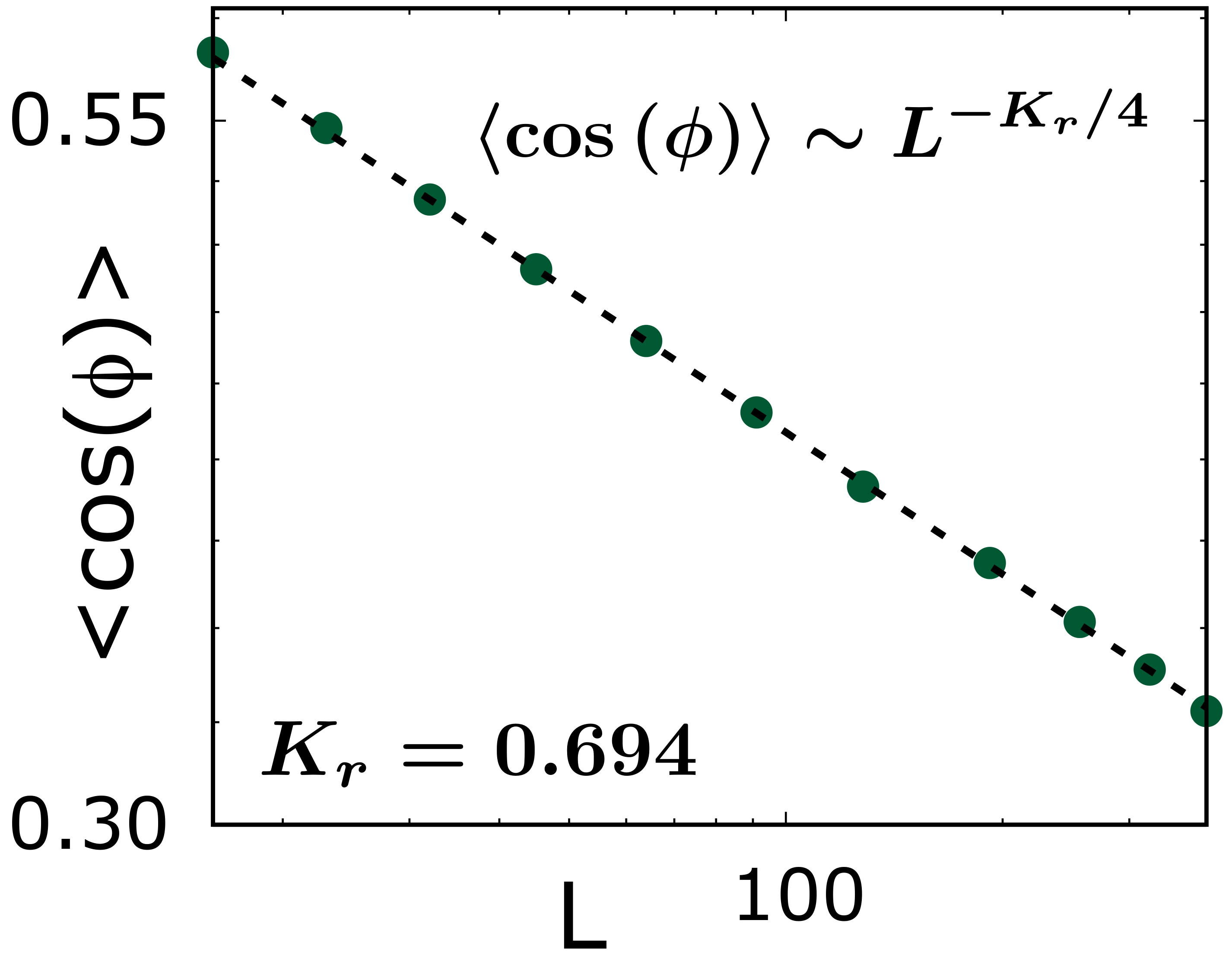}
\includegraphics[width=0.33\linewidth]{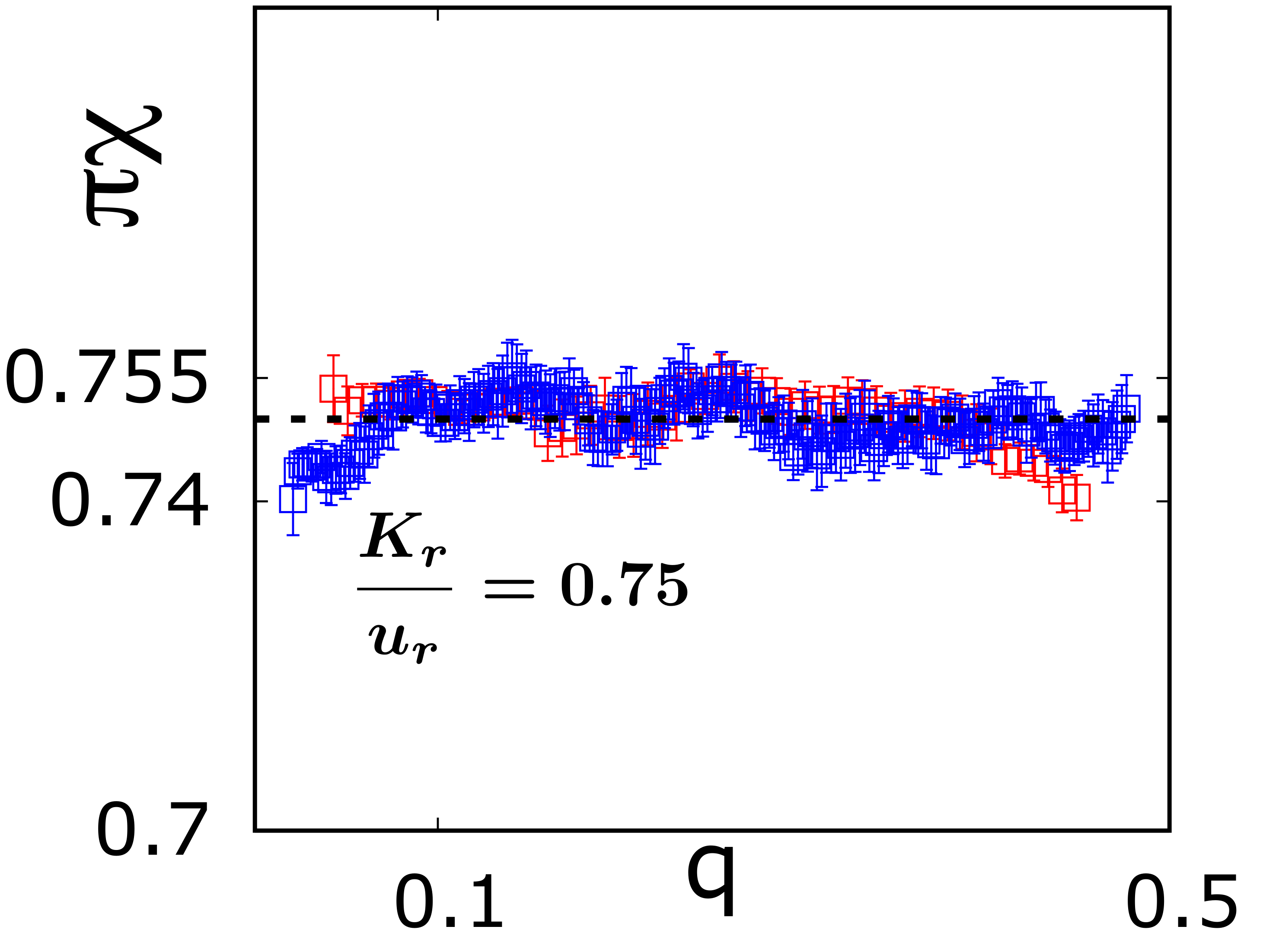}
\includegraphics[width=0.32\linewidth]{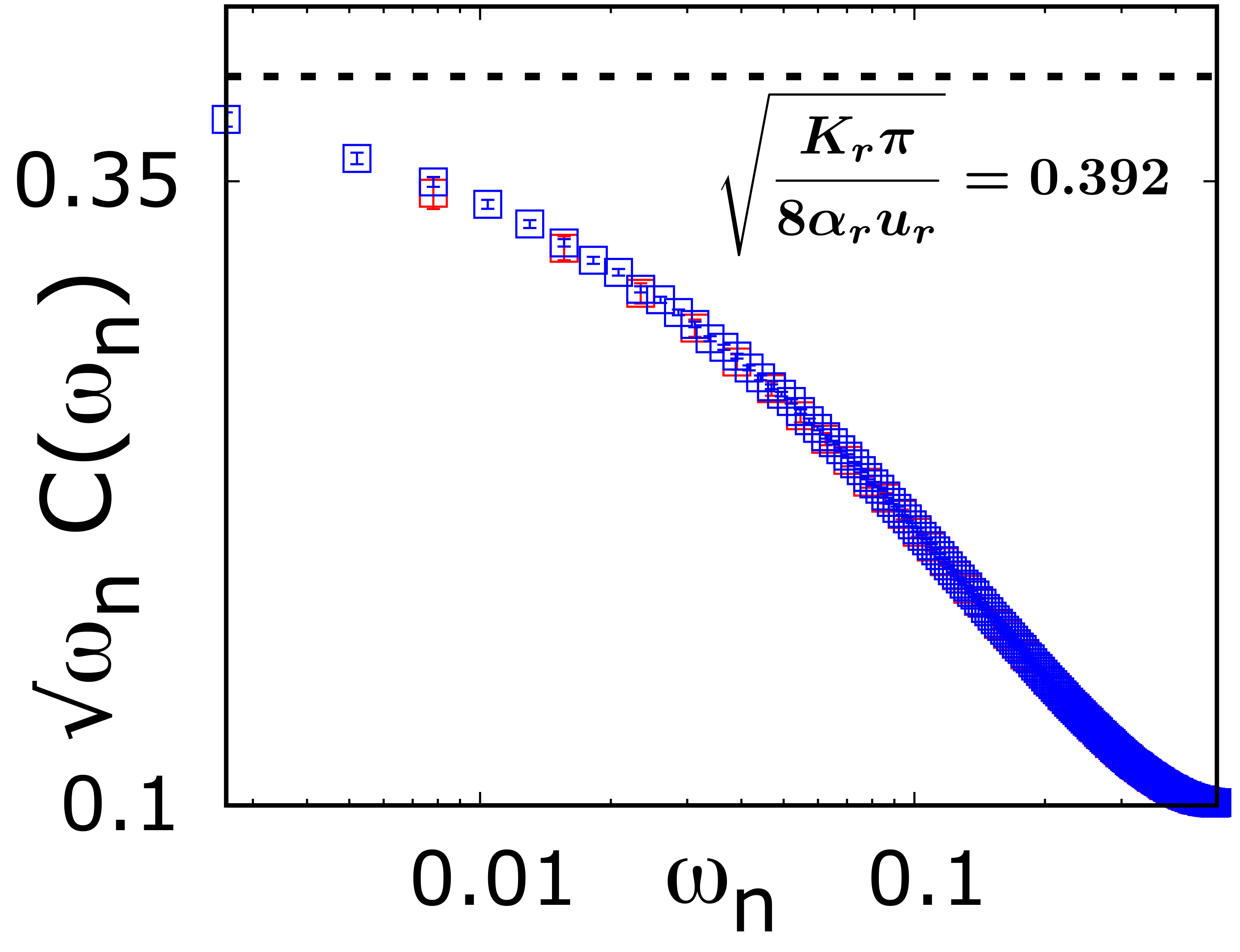}
\includegraphics[width=0.31\linewidth]{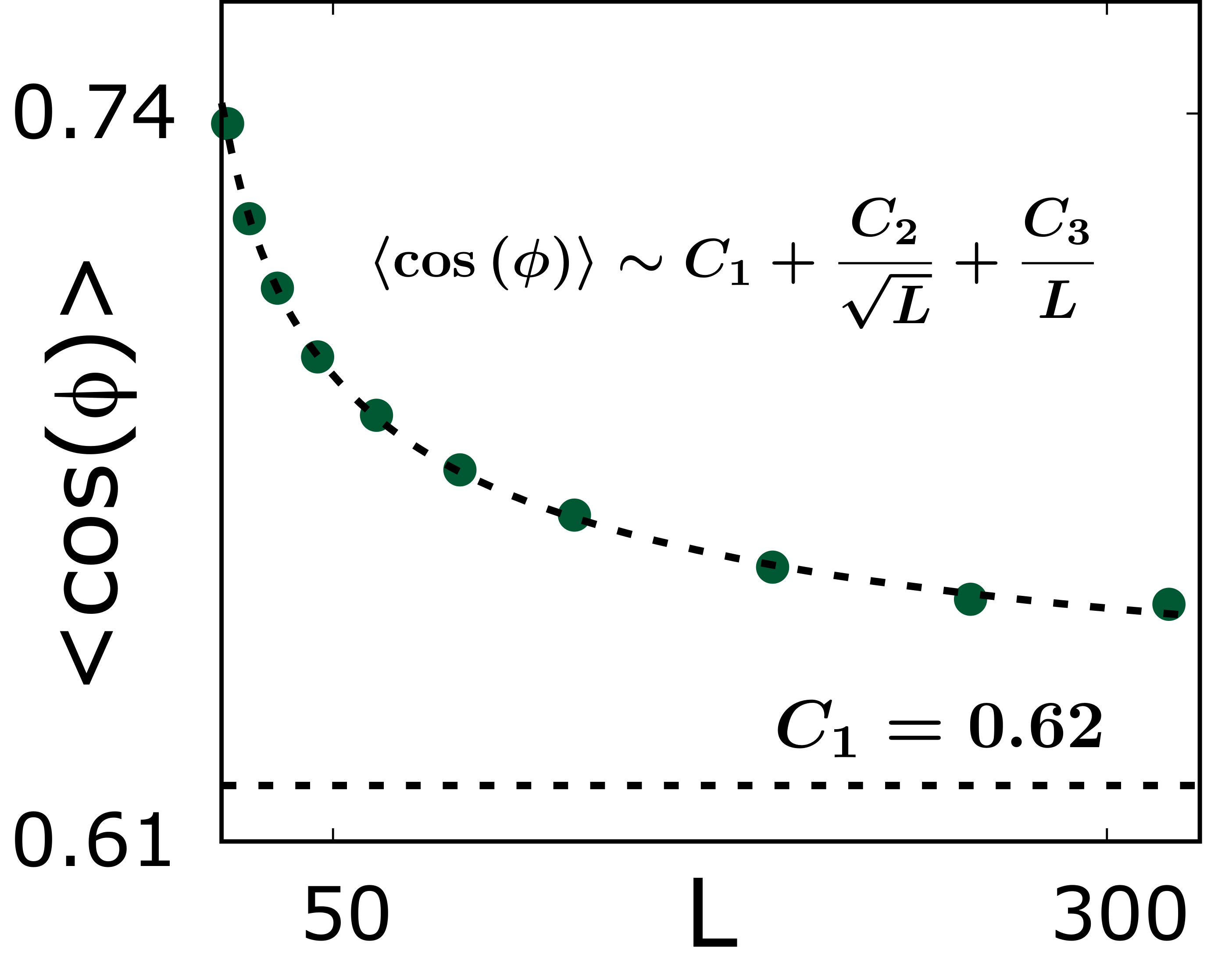}
\caption{Calculation of different quantities for $K=0.75$ that characterizes LL ($\alpha=1$, top row) and dissipative phase (  $\alpha=8$, bottom row).  Blue and red points correspond to $L=\beta=384$ and $L=\beta=128$, respectively. \textit{Left} : Due to symmetry, $\pi \chi = K_r/u_r$ is equal to $K/u=0.75$ for all values of $\alpha$ and all lengthscales.  \textit{Middle} : For $\alpha=1$,  $\omega_n C(\omega_n)$ saturates to $K_r/2 = 0.349$ as $\omega_n \to 0$; whereas for $\alpha=8$,  $\sqrt{\omega_n} C(\omega_n)$ saturates to  $ [K_r \pi/ (8 \alpha_r u_r)]^{1/2} =0.392$.  The other fitting constants are $a_1 = 0.2493$ and $a_2 = 3.572$.  \textit{Right} : For $ \alpha=1$, $ \langle \cos(\phi) \rangle$ decays as a power law, which allows us to extract $K_r=0.694$, consistent with the fit of $\omega_n C(\omega_n)$. For $\alpha=8$ it saturates to a constant, as predicted  by the variational ansatz (the  fit gives $c_1 = 0.62,$ $c_2 = 0.603$ and $c_3 = -0.531$). }
\label{fig:orderpar} 
\end{figure*}

The bath introduces a long-range cosine interaction in $\tau$ direction only and the strength of this potential is controlled by the parameter $\alpha$.  A perturbative RG study \cite{cazaillashort} shows that for $K<K_c=0.5$, the cosine term is relevant and the LL phase is destroyed, whereas for $K>K_c$ and small $\alpha$,  the system stays in LL phase but with renormalised LL parameters $K_r$ and $u_r$.  For $K \gtrsim K_c$, the transition is of KT type:  The critical point $\alpha_c(K)$ is still LL with $K_r=K_c=0.5$.  The nature of the dissipative phase is not clear : For moderate $K$ and very large $\alpha$, the action should be gapless and harmonic, obtained by the quadratic expansion of the cosine term.  For $K \gg K_c$, a large-N argument suggests the existence of a gapped disordered phase.  Monte-Carlo simulations  \cite{BIBL}  were performed on the 1D hard-core bosonic chain, which can be mapped to free fermions ($K=1$). Increasing $\alpha$, they found that $\chi$ increases and at $\alpha_c$, the system undergoes a continuous second-order phase transiton with vanishing $\rho_s$.  Below, we propose a simple scenario able to conciliate the puzzle of contradictory results.

\textit{Methods} : To make progress, we focus on the interacting bosonised action. On one side, we compute the correlation functions numerically by generating equilibrated configurations from the action with the help of Langevin dynamics.  On the other hand, to understand the properties of the dissipative phase, we improve the harmonic expansion proposed in \cite{cazaillashort} using a variational approach where we obtain an effective quadratic action by minimizing the variational free energy (\cite{Note1}, Sec. 2).  The propagator of this action is given by:
\begin{align*}
G^{-1}_{\text{var}}(q,\omega_n) = \frac{u_rq^2}{2 \pi K_r}+ \frac{\alpha_r}{\pi^2}\left |\omega_n\right| + a_1\left|\omega_n \right|^{\frac{3}{2}} +a_2\omega_n^2  \tag{7}
\label{eq:var_ans}
\end{align*}
The macroscopic behaviour of this phase depends only on the two parameters $u_r/K_r $ and $\alpha_r$.  The parameters $a_1$ and $a_2$ are introduced to account for finite size effects. From the analysis of our result, we will show that by varying $\alpha$, the long-distance properties are always captured either by the LL or by the variational propagator with renormalized parameters $u_r, K_r$ and $\alpha_r$. 

\begin{figure}[h]
\includegraphics[width=0.45\textwidth]{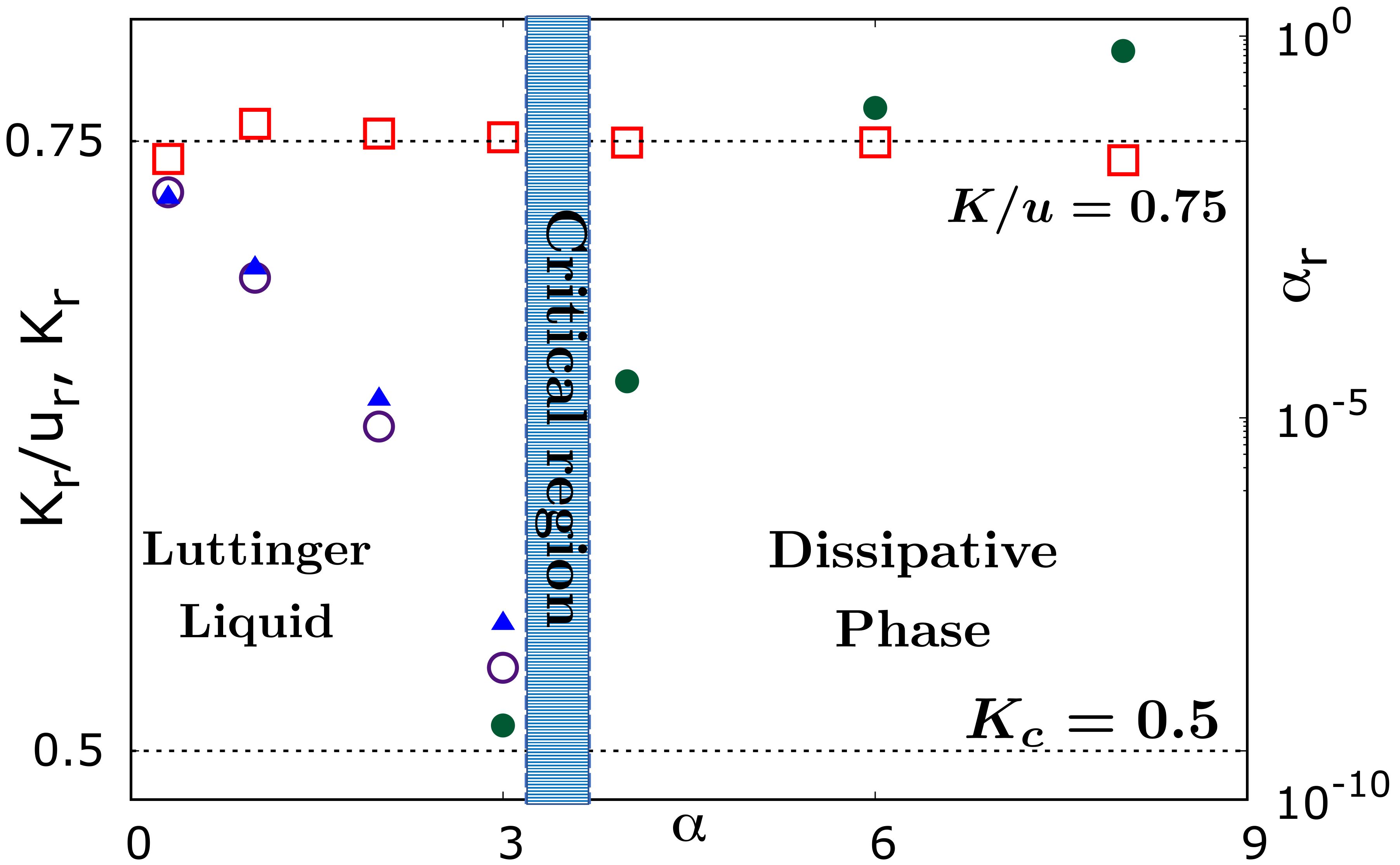}
\caption{Behaviour of different renormalised parameters as a function of dissipative coupling $\alpha$.  $K_r/u_r$ (red square points) remains constant and equal to $K/u = 0.75$ for all values of $\alpha$.  $K_r$, extracted from $\langle \cos (\phi)\rangle$ analysis (purple hollow circular points) agrees with the one from the $C(\omega)$ analysis (blue solid triangular points). It approaches $K_c=0.5$ as $\alpha$ reaches the critical point. The parameter $\alpha_r$ (green solid circular points) starts to be defined in the  dissipative phase and increases rapidly with increase in $\alpha$.  This behaviour of the parameter allows to locate the different phases : For $\alpha \in [0,3]$, the system is in LL phase, whereas for $\alpha \in [4,8]$ the system is in dissipative phase. The phase transition takes place for $\alpha \in (3,4)$.    }
\label{fig:phase} 
\end{figure}

\textit{Results} : In the following, we present our results for the correlation functions of the action $S=S_{\text{int}}+S_{\text{LL}} $ with $u=1,K=0.75$ and different $\alpha$. For our simulations, we set $\beta=L$. The first observation is that the action of Eq.(\ref{eq:S_int}) is invariant under tilt transformation(\cite{Note1}, Sec.  5). As a consequence, $\chi$ is not affected by the presence of $S_{\text{int}}$. We measure $K_r/u_r$ both at low and high $\alpha$ as shown in fig.  \ref{fig:orderpar} left. Note that the susceptibility corresponds to $q \to 0$ limit, but due to the symmetry, $K_r/u_r$ is invariant at all length scales and all values of $\alpha$.  We conclude that $K_r/u_r = K/u $ for all values of $\alpha$.  In principle, we should now measure the stiffness, but if we focus only on the limit $q \to 0$, it strongly fluctuates. Instead we introduce a function $C(\omega_n) =(1/\pi L) \sum_q \left| \phi(q,\omega_n) \right|^2$, which for small $\omega_n$ behaves as $K_r/2\omega_n$ in the LL phase and as $\sqrt{K_r/8\pi u_r}(\alpha_r\omega_n/\pi^2+a_1 \omega_n^{3/2}+a_2 \omega_n^2)^{-1/2}$ according to the variational prediction. From fig. \ref {fig:orderpar} middle, we see that indeed for small $\alpha$,  $C(\omega_n)$ behaves as expected for the LL phase while for large $\alpha$,  $C(\omega_n)$ shows an agreement with the variational approach. To confirm our prediction, we compute an independent quantity, namely $\langle \cos(\phi) \rangle$.  Due to the fluctuating nature of the bosonised field in LL phase, this quantity goes to 0 with a finite size behaviour $\langle \cos(\phi) \rangle_{\text{LL}} \sim L^{-K_r/4} $. On the other hand, in the dissipative phase the cosine potential pins the field. Using the variational ansatz we predict $ \langle \cos(\phi) \rangle_{\text{var}} \sim c_1 + c_2/\sqrt{\beta} + c_3/\beta $, where $c_1,c_2,c_3$ are fitting cut-off dependent parameters(\cite{Note1}, Sec.  3) . Fig. \ref {fig:orderpar} right confirms the scenario of a transition between an LL to a dissipative phase described by the variational ansatz.  Moreover, the value of $K_r$ extracted from $\langle \cos(\phi) \rangle$ matches nicely with the prediction of $C(\omega_n)$.  In Fig.  \ref{fig:phase}, we rationalise our results of the renormalized parameters, obtained by varying $\alpha$. For $K=0.75$, the critical region where the stiffness vanishes is $\alpha \in (3,4) $. Moreover just before the transition, $K_r$ approaches 0.5, as predicted by KT transition.  In (\cite{Note1}, Sec.  6), we provide further results for $u=1,K=0.55$, which are also in agreement with this picture.

\textit{Transport properties}: With our approach, one can compute thermodynamic quantities and does not have direct access to transport properties. However, via Wick rotation, the conductivity can be related to the propagator :
\begin{equation*}
\sigma(\omega) = \frac{e^2}{\pi^2 \hbar}  \left[\omega_n G(q=0,\omega_n)\right]_{i\omega_n \to \omega + i \epsilon} \tag{8}
\end{equation*}
For Luttinger liquid,  the DC conductivity $\sigma_{\text{DC}} \equiv \text{Re} \left[ \sigma(\omega \to 0)\right] =(e^2 uK/\hbar) \delta(\omega) $,  which shows the system is perfectly conducting. For the dissipative phase, after Wick rotation,we get $\sigma_{\text{DC}}=e^2/\hbar\alpha_r$,  proving that the system has finite conductivity.  For a generic bath, $G(q=0,\omega_n) \sim 1/(\alpha_r|\omega_n|^s)$,  and hence $\text{Re}[\sigma(\omega)]=(e^2/\hbar\alpha_r)(\epsilon/(\omega^2+\epsilon^2)^{s/2})$. Specially when the bath is subohmic ($s<1$), the DC conductivity of the system goes to 0, which is a signature of bath induced localization in the system.

\textit{Discussion}: It remains important to clarify how to conciliate our observations of a KT transition with $K_c=0.5$ and the hardcore bosonsic Monte Carlo simulations (at $K=1$ instead of $K=0.75$) that show a vanishing stiffness at the transition \cite{BIBL}.  A possibility is that it is an artifact of the commensurate-incommensurate crossover of the system as the system size as well as the incommensurate parameter used in the Monte-carlo study is small.   Another possibility remains that our action misses some term that is relevant for the microscopic lattice model.

\textit{Conclusion}: On a more general framework,  Many efforts are currently being made to observe phase transition in open quantum systems.  Often the physical bath is replaced by quantum measurents.  For a single degree of freedom, this gives rise to the Zeno effect \cite{SudarshanZeno,WayneZeno} where the spin dyamics is completely frozen for very frequent measurements. To observe similar phenomena in many-body systems,  solving the Lindblad equation is a very commonly used formalism \cite{Thibaud}.  However, with this approach the numerics are often limited to very small system sizes. Another popular method is to approximate these systems via quantum circuits through which people have been able to show phase transitions \cite{CircuitZeno1,CircuitZeno2,CircuitZeno3,
CircuitZeno4,CircuitZeno5,CircuitZeno6,CircuitZeno7,CircuitZeno8,
CircuitZeno9,CircuitZeno10}, but these models are very simplistic and doesn't capture the physics fully. Our work shows that localisation effects can be induced by optical phonons, provided that the nature of the bath is slow enough (ohmic or subohmic). 
 




 

\begin{acknowledgments}

{\it Acknowledgments:} 
This work is supported by `Investissements d'Avenir' LabEx PALM
(ANR-10-LABX-0039-PALM) (EquiDystant project, L. Foini) and is supported in part by the Swiss National Science Foundation under Division II. We thank T. Maimbourg and V. Schimmenti for helpful discussions. This work was performed using HPC/AI resources from GENCI-IDRIS (Grant
2022-AD011013581) and GENCI-TGCC (Grant
2022-AD011013555) .
\end{acknowledgments}

\appendix


\bibliography{references}

\onecolumngrid
\newpage

\begin{center}
{\Large Supplementary Material \\ 
}
\end{center}
\section{Bosonisation of the XXZ spin chain with dissipative bath}
In this section, we apply Bosonisation on the one-dimensional XXZ spin chain coupled with dissipative baths, arriving at an equivalent field theory.  The hamiltonian of the system is given by :
\begin{align}
\begin{split}
H &= H_{\text{S}}+H_{\text{B}}+H_{\text{SB}} \\
H_{\text{S}} &= \sum_{j=1}^N J_z S_j^z S_{j+1}^z + J_{xy} \left(S_j^xS_{j+1}^x + S_j^yS_{j+1}^y  \right)\\
H_{\text{B}} &= \sum_{j k} \frac{P^2_{j k}}{2 m_k} + \frac{m_k \Omega_k^2}{2} X_{kj}^2\\
H_{\text{SB}} &= \sum_{j=1}^N S_j^z \sum_k \lambda_k X_{k j}
\end{split}
\end{align}
Here we focus on the equilibrium properties at the inverse temperature $\beta=\infty$.  To tackle the dissipative problem, there are two different approaches. One can map to an equivalent fermionic system via Jordan-Wigner transformation and apply bosonisation to arrive at a two-dimensional field theory.  Alternatively, the quantum hamiltonian is mapped onto hard-core bosonic system via Holstein-Primakoff transformation and then numerically simulated via Quantum Monte-Carlo methods (Indeed Bosons don't suffer from Sign problem). In both cases, to integrate the bath degrees of freedom, one has to introduce the path integral description of the system. The action associated to the bath and the interaction between the bath and the system is identical for bosons and fermions and is given by:
\begin{equation}
S_{\text{B}}+S_{\text{SB}} =\int_0^{\beta} d\tau \sum_{j=1}^N \left[ \left(n_j(\tau) -\frac{1}{2}\right)\sum_k \lambda_{\alpha} X_{k j} +  \sum_{k}\left(m_k \dot{X}^2_{k j} + \frac{m_k \Omega^2_{k}}{2} X^2_{k j} \right) \right]
\end{equation}
Where $n_j = S^z_j + \frac{1}{2}$ is the density operator.  Note that we have introduced the imaginary time $\tau \in (0,\beta)$. Now, we can integrate out the bath degrees of freedom and arrive at an effective action for the system degrees of freedom only,  where the effect of the bath is encoded in the interacting part $ S_{\text{int}}$:  
\begin{equation}
S_{\text{int}} = -\iint_0^{\beta} d\tau d\tau' \sum_{j=1}^N \left[n_j(\tau)-\frac{1}{2}\right]D(\tau-\tau')\left[n_j(\tau')-\frac{1}{2}\right] 
\label{eq:actionNB}
\end{equation}
Here $D(\tau-\tau') $ is the dissipative kernel which is produced from integrating over the bath modes. In particular, its Fourier transform can be expressed in terms of the bath spectral function $J(\Omega)$ (defined in the main text):   
\begin{equation}
D(\omega_n) =\frac{2}{\pi} \int_0^{\infty} J(\Omega) \frac{\Omega}{\omega_n^2 + \Omega^2}
\end{equation} 
Using the form $J(\Omega) = \pi \alpha \Omega$ ($s=1$), we get $D(\tau-\tau') \sim \alpha|\tau-\tau'|^{-2}$.  \\
 
The Bosonisation procedure of the XXZ spin chain is well known \cite{giamarchibook}. Away from the sector of zero magnetisation, one recovers the action for the well-known Luttinger liquid model $ S_{\text{LL}} = \frac{1}{2 \pi} \int_0^L dx \int_0^{\beta} d\tau \left[ \frac{1}{u K} (\partial_{\tau} \phi(x,\tau))^2 + \frac{u}{K} (\partial_{x} \phi(x,\tau))^2 \right]$. At zero magnetisation, there is an extra term in the action $S_{\text{cos}} = -\frac{J_z}{2 \pi^2} \int_0^L dx \int_0^{\beta} d\tau \cos (4\phi(x,\tau))$, which is irrelevant for $K>1/2$. The constants $u$ and $K$ are called Luttinger liquid parameters and they depend on $J_{xy}$ and $J_z$.  These parameters can be exactly calculated from Bethe Ansatz (eg. $ K^{-1}_{\text{Bethe ansatz}} = (2/\pi) \arccos (-J_z/J_{xy})$) and they match with the Bosonisation prediction in the regime $J_z <<J_{xy} $ ($K^{-1}_{\text{bosonisation}}=\sqrt{1+4J_z/\pi J_{xy}}$).  However, away from half-filling (non-zero magnetisation sector),  the bosonisation prediction between the LL parameters and the spin chain  are slightly more complicated and given by $uK = a J_{xy} \sin (q_F a)$ and $u/K = uK(1+ \frac{2 a J_z}{\pi v_F}\left[1-\cos(2 q_F a) \right])$, where $q_F$ is the Fermi momentum of the spin chain and $a$ is the lattice spacing kept for dimensional matching reasons.\\

To bosonise Eq. (\ref{eq:actionNB}), we recall that the bosonized version of $S_j^z$ is given by : 
\begin{equation}
\hat{S}^z = - \frac{1}{\pi} \nabla \phi + \frac{1}{\pi a} \cos (2 \phi - 2 q_F x)
\end{equation}
Using Eq. (\ref{eq:actionNB}),  the dissipative part of the action is given by:
\begin{align}
\begin{split}
S_{\text{int}} &= -\frac{1}{2 \pi^2} \int_0^L dx\iint_0^{\beta} d\tau d\tau'\left(- \nabla \phi(x,\tau) + \frac{1}{a} \cos (2 \phi (x,\tau) - 2 q_F x) \right)\\
&\times D(\tau-\tau')\left(- \nabla \phi(x,\tau') + \frac{1}{a} \cos (2 \phi (x,\tau') - 2 q_F x) \right)
\end{split}
\end{align}
After multiplying all the terms, one can put $a=1$, which was there for dimensional purposes. For the expansion, we will be making a few observations here: 
\begin{itemize}
\item At equal time ($\tau=\tau'$), the dissipative action is $(S_j^z)^2$, which is identity.  Hence, this term doesn't contribute anything to the physics and we can neglect the constant term in $ D(\tau-\tau')$. 

\item the terms with of the form $\nabla \phi(\tau) \cos(2\phi(\tau')-2q_F x)$ will oscillate rapidly for non-zero magnetisation due to the $2 q_F x$ term and hence it will integrate to 0. Similarly, the $\cos(2\phi(\tau)-2q_F x) \cos(2\phi(\tau')-2q_F x)$ term can be broken up in two terms; one of these terms will be of the form $\cos(2(\phi(\tau)+\phi(\tau'))-4q_F x)$. This term can also be integrated to 0 due to the rapidly oscillating term $4 q_F x$.

\item the $\nabla\phi(\tau)\nabla\phi(\tau')$ term is the forward scattering term and is irrelevant by power counting.

\end{itemize}
Hence, the action of the full system turns out to be:
\begin{align}
\begin{split}
S = S_{\text{LL}}- \frac{\alpha}{4 \pi^2} \int_0^L dx \iint_0^{\beta} d\tau d\tau' \frac{\cos(2\phi(x,\tau)-2\phi(x,\tau'))}{|\tau-\tau'|^2} 
\end{split}
\end{align}

\section{Variational ansatz}
In this section, we show a detailed calculation of our variational ansatz to find an effective action that captures the effect of the dissipative term \cite{feynmann}.

\vspace{5px}
The full bosonised action of the system is given by $S = S_{\text{LL}}+S_{\text{int}} $ :
\begin{align}
\begin{split}
S_{\text{LL}} &=\frac{1}{2\pi K} \int dx d\tau \left[ \frac{1}{u} \left(\partial_{\tau} \phi(x,\tau) \right)^2 + u\left(\partial_x \phi(x,\tau) \right)^2  \right] \\
 S_{\text{int}} &= - \frac{\alpha}{4\pi^2} \int dx d\tau d\tau' \frac{ \cos \left(2\left(\phi(x,\tau)-\phi(x,\tau') \right)\right)}{|\tau-\tau'|^{2}}
\end{split}
\end{align}

We would like to find an effective quadratic action of the form  $S_{\text{var}} = \frac{1}{2\beta L} \sum_{q,\omega_n} \phi^*(q,\omega_n)G^{-1}_{\text{var}}(q,\omega_n) \phi(q,\omega_n)$.  We use the variational method : we minimise the free energy $ F_{\text{var}} = -\frac{1}{\beta} \sum_{q,\omega_n} \log G_{\text{var}} + \frac{1}{\beta} \langle S-S_{\text{var}} \rangle_{S_\text{var}} $ (with $(S-S_{\text{var}})$ averaged over $S_{\text{var}}$) with respect to the variational Green's function :
\begin{equation}
\begin{gathered}
G^{-1}_{\text{var}}  =  \frac{1}{2 \pi K}\left(uq^2+\frac{\omega_n^2}{u} \right) + \frac{\alpha}{\pi^2}\int d\tau D(\tau) \left(1-\cos \omega \tau \right)e^{ \left(- \frac{1}{\pi^2} \int_{-\infty}^{\infty} dq d\omega \ G_{\text{var}}(q,\omega) \left(1-\cos \omega\tau \right) \right)}
\end{gathered}
\end{equation}
We try to solve this self-consistent equation making the following ansatz : $ G^{-1}_{\text{var}}(q,\omega) = \frac{1}{2 \pi K}\left(uq^2+\frac{\omega_n^2}{u} \right) + \frac{\alpha}{\pi^2} F(\omega)$, where $F(\omega) = a(\alpha)\left| \omega \right|^{\psi} +  b(\alpha) \left| \omega \right| $. With this assumption, for large $\tau$, the behaviour of $G(\omega)$ is governed by the $|\omega|$ term.  It can be easily shown that $\int_{\infty}^{\infty} dq d\omega \ G_{\text{var}}(q,\omega) \left(1-\cos \omega\tau \right) \approx C(\alpha) - \left(\frac{\tau_c(\alpha)}{\tau}\right)^{1/2}$,  where $C(\alpha)$ and $\tau_c(\alpha)$ are $\alpha$ dependent constants. For a more systematic expansion in powers of $1/\tau$ one can use the results in \cite{Santachiara}. Putting this back into the self-consistent equation for $F(\omega)$ we obtain :
\begin{equation}
\begin{split}
a(\alpha)\left| \omega \right|^{\psi} +  b(\alpha) \left| \omega \right| &= \int d\tau D(\tau)  \left( 1 -\cos \omega \tau \right) e^{-\left[C(\alpha) - \left(\frac{\tau_c(\alpha)}{\tau}\right)^{1/2} \right]}\\
& \stackrel{\text{large} \ \tau}{\approx} \int d\tau D(\tau)  \left( 1 -\cos \omega \tau \right)e^{-C(\alpha)} \left( 1 +\left( \frac{\tau_c(\alpha)}{\tau}\right)^{1/2}\right)
\end{split}
\end{equation}
The $\omega$ dependence can be easily extracted from these equations, which turns out to be $|\omega|$ and $|\omega|^{3/2}$.  The coefficient of $|\omega|$ should be determined self-consistently and in our analysis we take it as a fitting parameter $\alpha_r/\pi^2$.  The coefficient in front of $\omega^2$ will be renormalised by higher order terms from variational analysis.  Hence, the variational propagator is given at low order in $\omega$ by $G^{-1}_{\text{var}} =  \frac{u}{2 \pi K}q^2+ \frac{\alpha_r}{\pi^2}\left |\omega_n\right| +a_1\left|\omega_n \right|^{3/2}+a_2 \omega_n^2$, where $a_1$ and $a_2$ are fitting coefficients.

\section{Calculation of system size dependence of Order Parameter}
In this section, we find an analytical expression for the quantity $\langle \cos (\phi(x,\tau) \rangle$, which is equal to $e^{-\frac{1}{2}\langle\left[ \phi(\vec{r}) \right]^2 \rangle}$ for Gaussian theories. 

For our variational ansatz, we need to calculate $ \langle \cos(\phi(\vec{r})) \rangle = \exp(S_1) = \exp \left( - \frac{1}{2 \beta L} \sum_{q,\omega_n} \frac{1}{\frac{u}{ \pi K}q^2+ \frac{2\alpha}{\pi^2}\left |\omega_n\right| +2a_1\left|\omega_n \right|^{3/2}+2a_2 \omega_n^2} \right)$.  In order to calculate the sum inside the exponential we send the limit of integration over $q$ from $0$ to infinity, and the integral over $\omega$ from $1/\beta$ to $1/l_0$, where $l_0$ is the microscopic cut-off.  By doing so, one can find the small $\omega$ behaviour of the sum as below :
\begin{equation}
\begin{gathered}
S_1 =  \frac{1}{4}\sqrt{\frac{K}{\pi u}} \left[ \frac{a_1 \pi^3}{(2 \alpha)^{\frac{3}{2}}l_0} - \frac{2 \pi}{\sqrt{2 \alpha l_0}} \right] + \sqrt{\frac{\pi K}{8 u \alpha}} \frac{1}{\sqrt{\beta}} - \frac{a_1 \pi^3}{4} \sqrt{\frac{K}{8 u \alpha^3}} \frac{1}{\beta}
\end{gathered}
\end{equation}

We put this back into the expression of $ \langle \cos(\phi(\vec{r})) \rangle$ and from a large $\beta$ (zero temperature limit) expansion, we obtain the finite size dependence of the order parameter:
\begin{align}
\begin{split}
\langle \cos(\phi(\vec{r})) \rangle_{\text{var}} &= c_1 + \frac{c_2}{\sqrt{\beta}} + \frac{c_3}{\beta}\\
c_1 &= \exp \left[ \frac{1}{4}\sqrt{\frac{K}{\pi u}} \left( \frac{a_1 \pi^3}{(2a)^{\frac{3}{2}}l_0} -\frac{2\pi}{\sqrt{2 \alpha l_0}}\right)\right]\\
c_2 &= c_1 \sqrt{\frac{\pi K}{8 u \alpha}}\\
c_3 &= c_1 \left( \frac{\pi K}{16 u \alpha} - \frac{a_1 \pi^3}{4} \frac{K}{8 u \alpha^3}\right)
\end{split}
\end{align}

\section{Numerical details}
In this section, we describe the numerical procedure for this work.  In a nutshell, we and denoted $\phi_{ij}$, where $i \in [1,L]$
and $j \in [1,\beta]$ with periodic boundary conditions in both directions. 
Our strategy is to start from a flat interface $\phi_{ij}=0$ at $t=0$ and then let it evolve according to the Langevin equation  \cite{StochasticRK2}:
\begin{equation}\label{Langevin}
\frac{d \phi_{ij}(t)}{dt} = - \frac{\delta S[\phi_{ij}(t)]}{\delta  \phi_{ij}} + \eta_{ij}(t)
\end{equation}
Where $\eta_{ij}(t)$ is a white noise, specified by the correlations $\langle \eta_{ij}(t) \rangle = 0$ and $\langle \eta_{ij}(t) \eta_{i'j'}(t')\rangle = 2 \delta_{i,i'}\delta_{j,j'}\delta(t-t')$.  Note that the time $t$ that appears in Eq. (\ref{Langevin}) should not be confused with the imaginary time $\tau$. When $t\to\infty$ the surface $\phi_{i,j}(t)$ obtained by direct integration of Eq. (\ref{Langevin}) is equilibrated with the action $S[\phi]$. Hence, the Langevin equation, which we numerically simulate, is given by :
\begin{align}
\begin{split}
\frac{d \phi_{ij}(t)}{dt} &= \frac{u}{K\pi}\left[\phi_{i+1,j}+ \phi_{i-1,j}-2\phi_{i,j}\right] +
\frac{1}{uK\pi}\left[\phi_{i,j+1} + \phi_{i,j-1}-2\phi_{i,j}\right] \\
&+ \frac{\alpha}{\pi^2} \sum_{j'}D(|j-j'|) \sin \left[ 2(\phi_{ij'}-\phi_{ij})\right] + \eta_{ij}(t)
\end{split}
\end{align}

In order to obtain a correct discretisation of the long-range kernel $D(j-j')$,we use the same protocol as in \cite{RossoKardar}. For $\beta \to \infty$, we set :
\begin{equation}
D(j-j') =  \int_0^{2\pi} \frac{d\omega}{2\pi} e^{i\omega(j-j')}(2(1-\cos(\omega)))^{1/2} = \frac{1}{(j-j')^2-\frac{1}{4}}
\end{equation} 
At finite $\beta$,  the periodic boundary condition are implemented as :
\begin{equation}
D(j-j') = \sum_{k=-\beta/2}^{\beta/2} \frac{1}{(|j-j'|+k\beta)^2-1/4} 
\end{equation}
To conclude, we remark that in the numerical integration the term $\frac{\delta S[\phi_{ij}(t)]}{\delta(\phi)}$ is multiplied by $\Delta t$, whereas $\eta_{ij}(t)$  by $\sqrt{\Delta t}$. Here we use the stochastic 2nd order Runge-Kutta algorithm for white noise  \cite{StochasticRK2}. Using this is preferable as this is much faster than the standard Euler's algorithm.  We choose the value of the Langevin time-step $\Delta t=0.05$. To benchmark the equilibration time of the surface, we used the harmonic approximation $\sin \left[ 2(\phi_{ij'}-\phi_{ij})\right] \to 2(\phi_{ij'}-\phi_{ij})$ that can be analytically solved. 

\section{Tilt Symmetry of the action}
In this section, we explain why the parameter $K/u$, which identifies with the susceptibility $\chi$,  remains constant for all dissipative coupling $\alpha$.  \cite{Elisabeth, Mesard}. To compute the susceptibility,  we introduce a finite magnetic field $h$ in the $z$-direction.  Then the susceptibility can be written as $\chi =\frac{\partial^2}{\partial (h\beta)^2}(\log Z[h])$,  where $Z$ is the partition function and $\beta$ is the inverse temperature of the system.  In the bosonised language,  the term $-h \sum_{j} S^z_j$ in the hamiltonian gives rise to the term $-\frac{h}{\pi}\int (\nabla \phi(x,\tau)) dx d\tau $ in the action.  Hence, the partition function of the system can be written as :
\begin{equation}
Z[h] = \int \mathcal{D}[\phi] \exp \left(-S_{\text{LL}}-S_{\text{int}}+\frac{h}{\pi}\int \nabla \phi(x,\tau) dx d\tau \right)
\end{equation}
One can rewrite the terms $\frac{u}{2 \pi K}(\nabla \phi)^2 - \frac{h}{\pi} \nabla \phi$ as $ \frac{u}{2 \pi K} (\nabla \phi - \frac{h K}{u})^2 - \frac{h^2 K}{2 \pi u}$. Introducting the tilt $\tilde{\phi} \to \phi- \frac{h K x}{u}$, the partition function can be re-written :
\begin{equation}
Z[h] = \int \mathcal{D}[\tilde{\phi}] \exp\left(-S_{\text{LL}}[\tilde{\phi}]-S_{\text{int}}[\tilde{\phi}]+ \frac{\beta^2 h^2 K}{2 u \pi}\right)
\end{equation}
The key point is that the interacting action $S_{\text{int}}$ is invariant under the tilt transformation $S_{\text{int}}(\tilde{\phi}+\frac{hkx}{u}) = S_{\text{int}}(\tilde{\phi}) $. From the previous equation, it can be easily seen that :
\begin{equation}
\log Z[h] =  \frac{\beta^2 h^2 K}{2 u \pi} + \log Z[h=0] 
\end{equation}
From this expression, the susceptibility can be easily computed, which is finally given by: \begin{equation}
\chi = \frac{\partial^2}{\partial (\beta h)^2}\frac{\beta^2 h^2 K}{2u \pi} = \frac{K}{u \pi}
\end{equation}
\newpage
\section{Results for $K=0.55$}
\begin{figure}[H]
\centering
\includegraphics[width=0.33\linewidth]{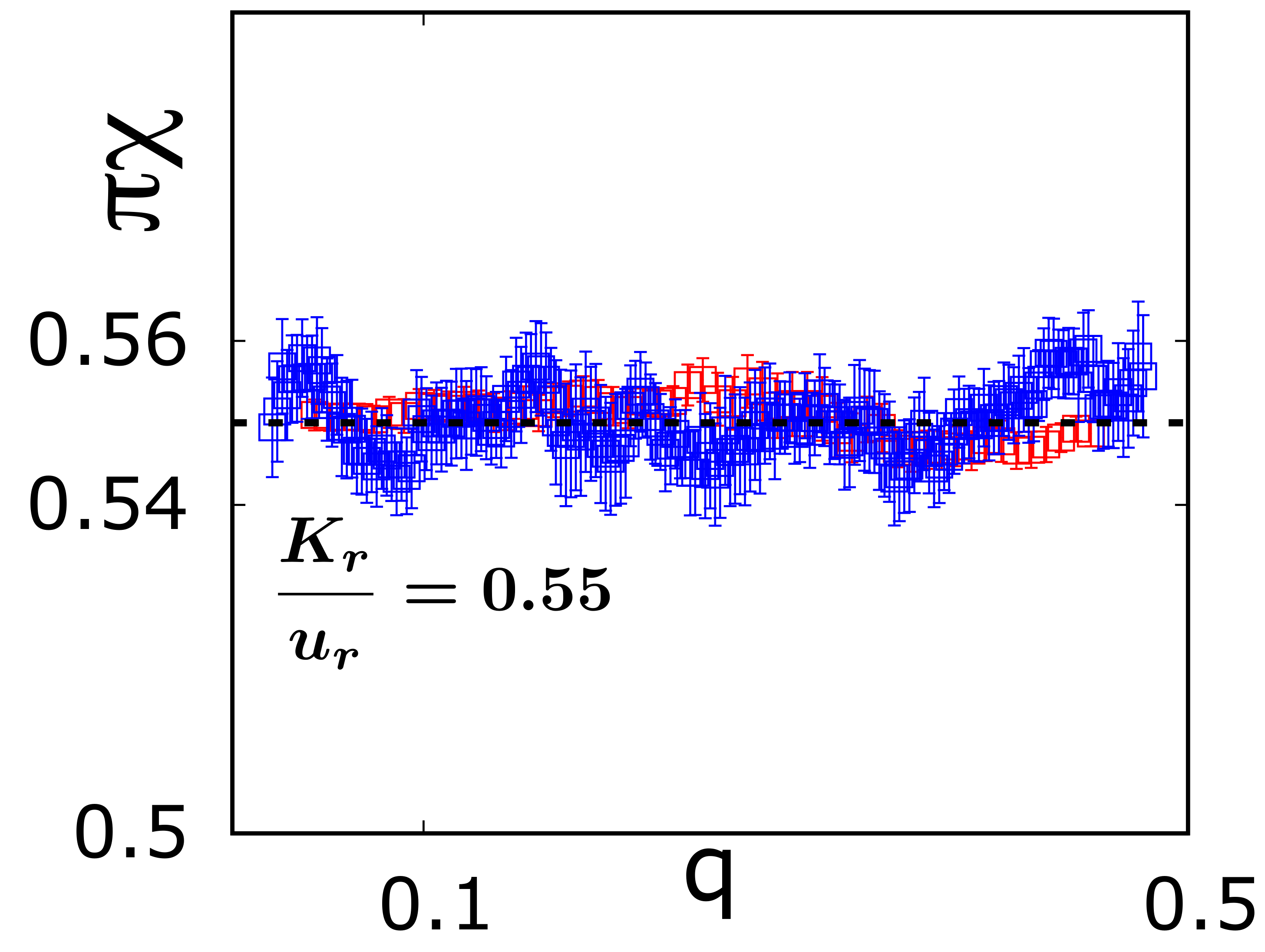}
\includegraphics[width=0.33\linewidth]{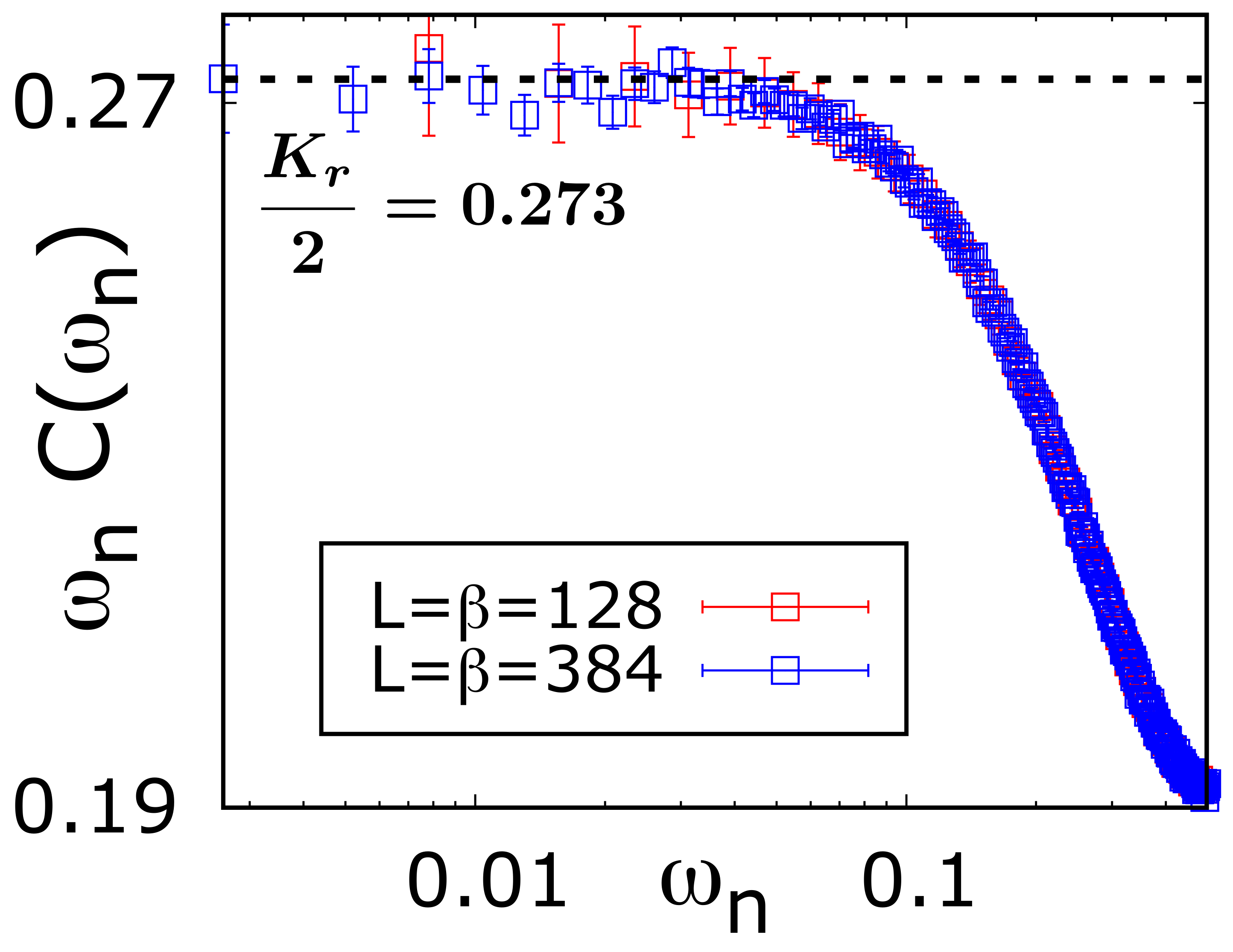}
\includegraphics[width=0.32\linewidth]{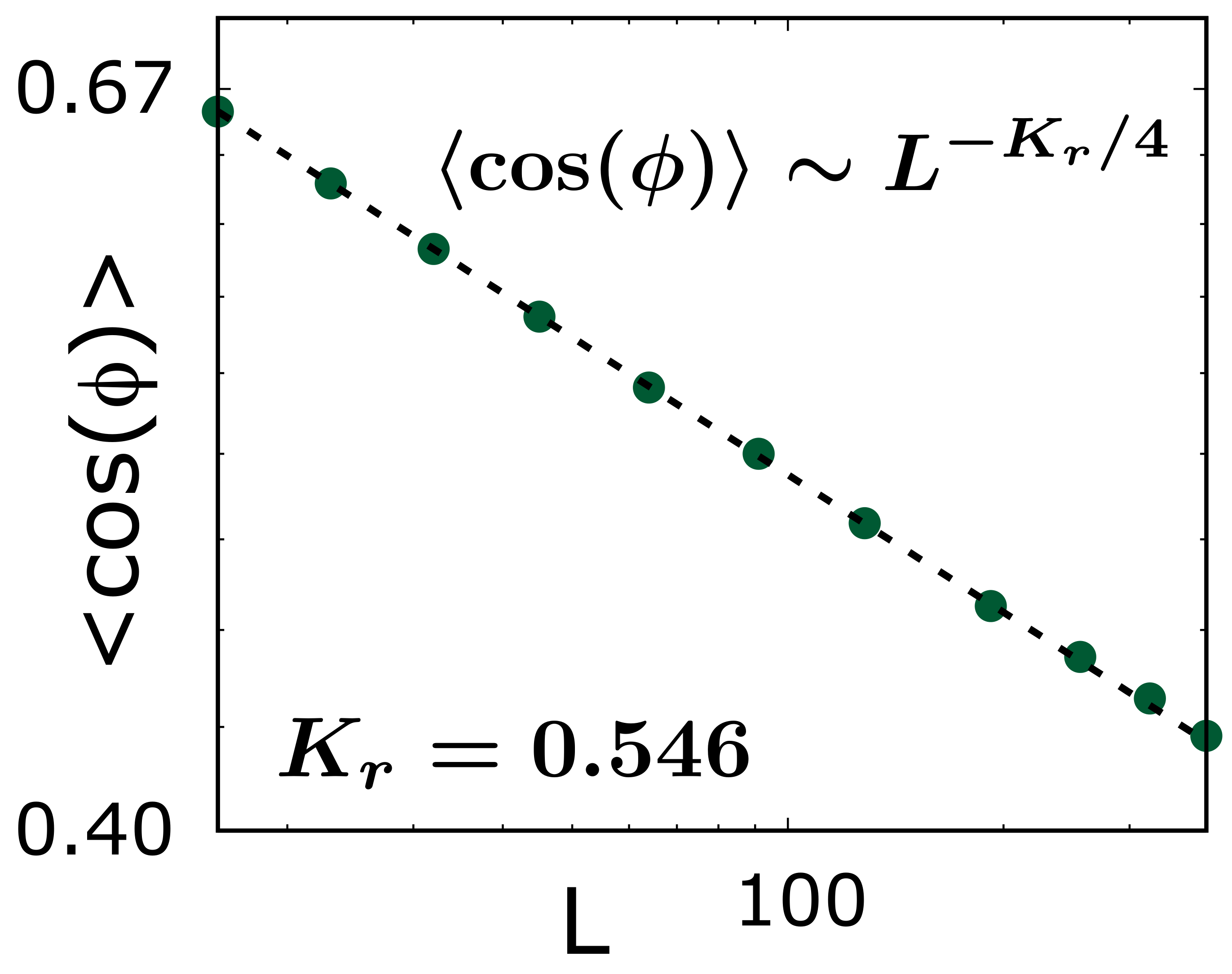}
\includegraphics[width=0.33\linewidth]{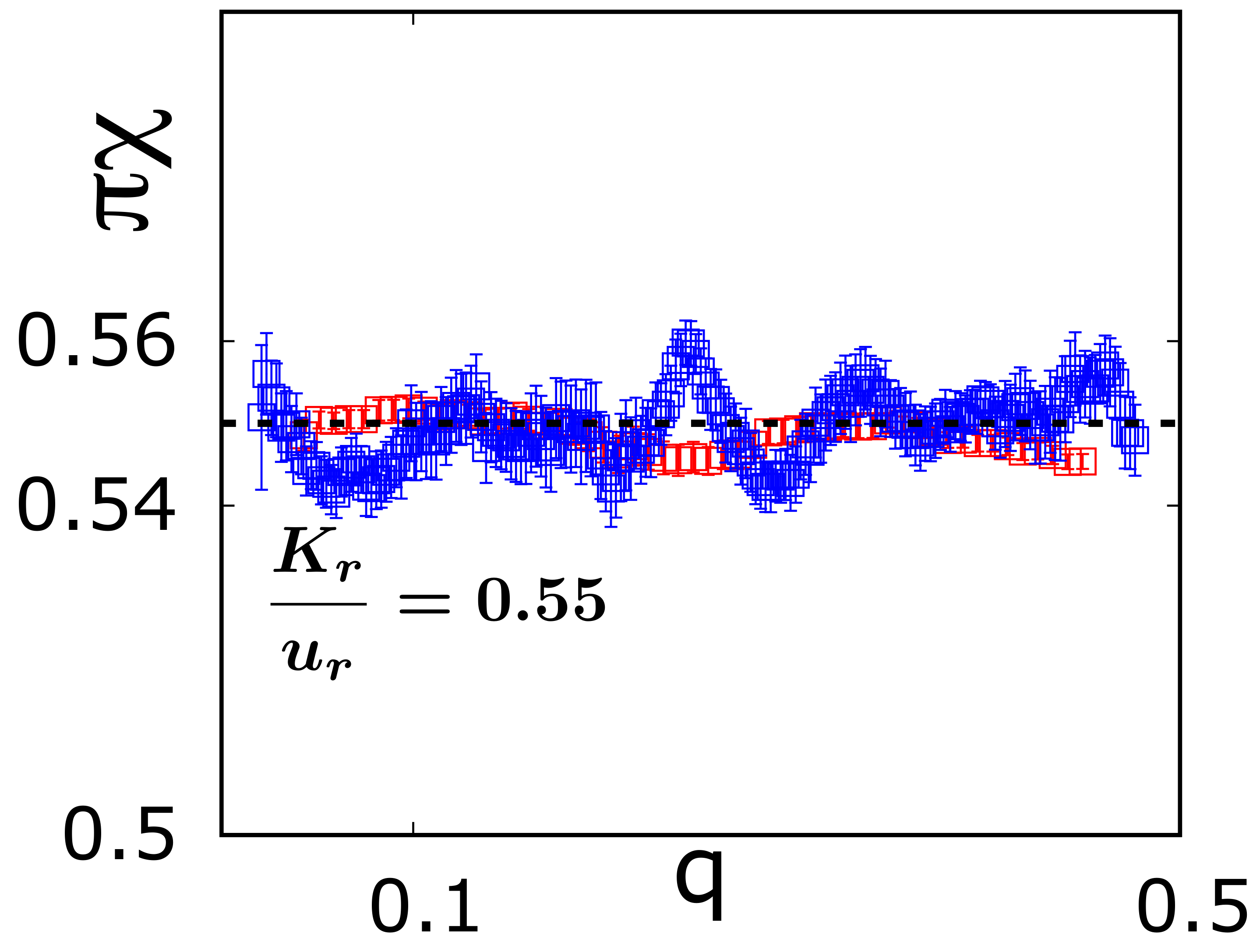}
\includegraphics[width=0.33\linewidth]{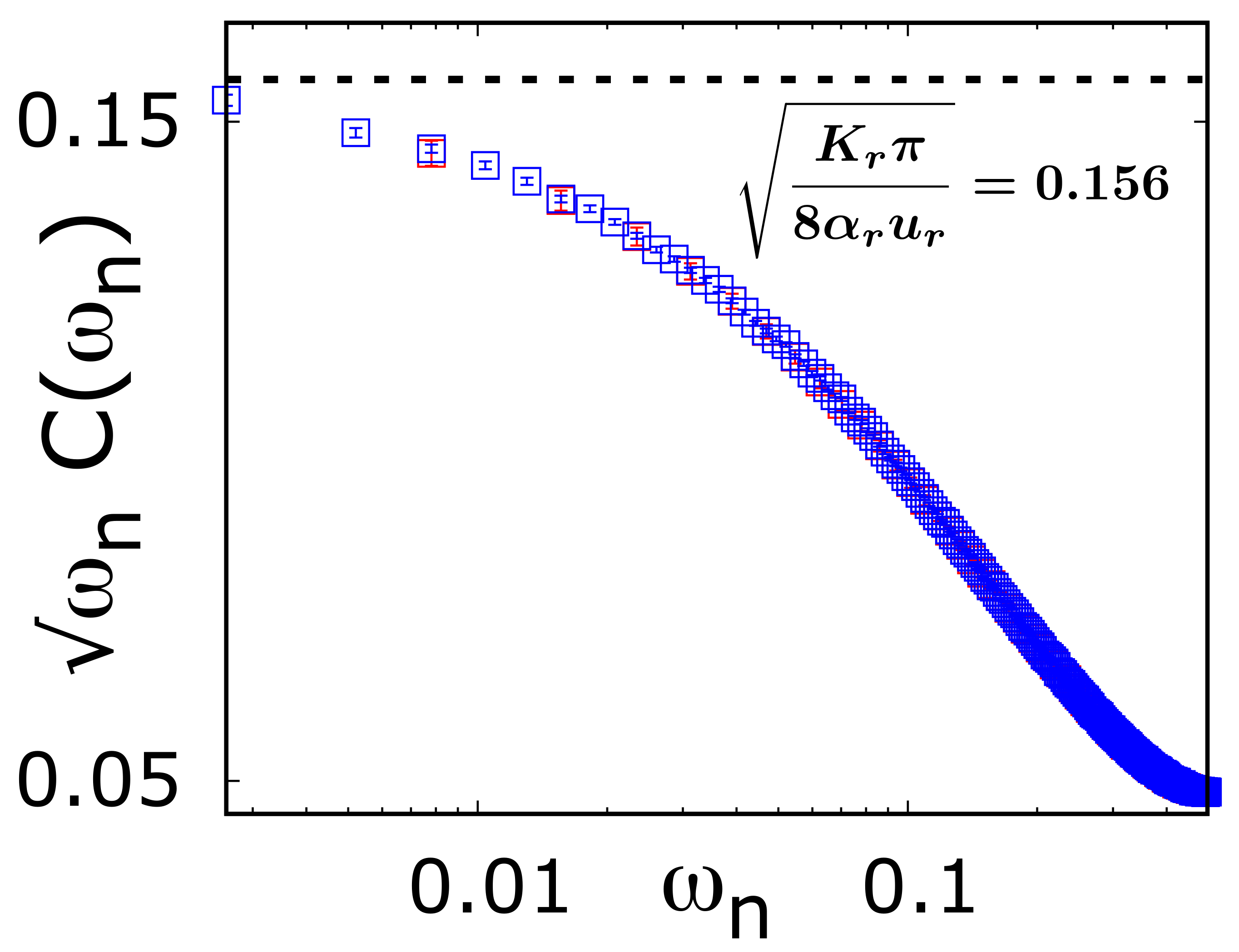}
\includegraphics[width=0.32\linewidth]{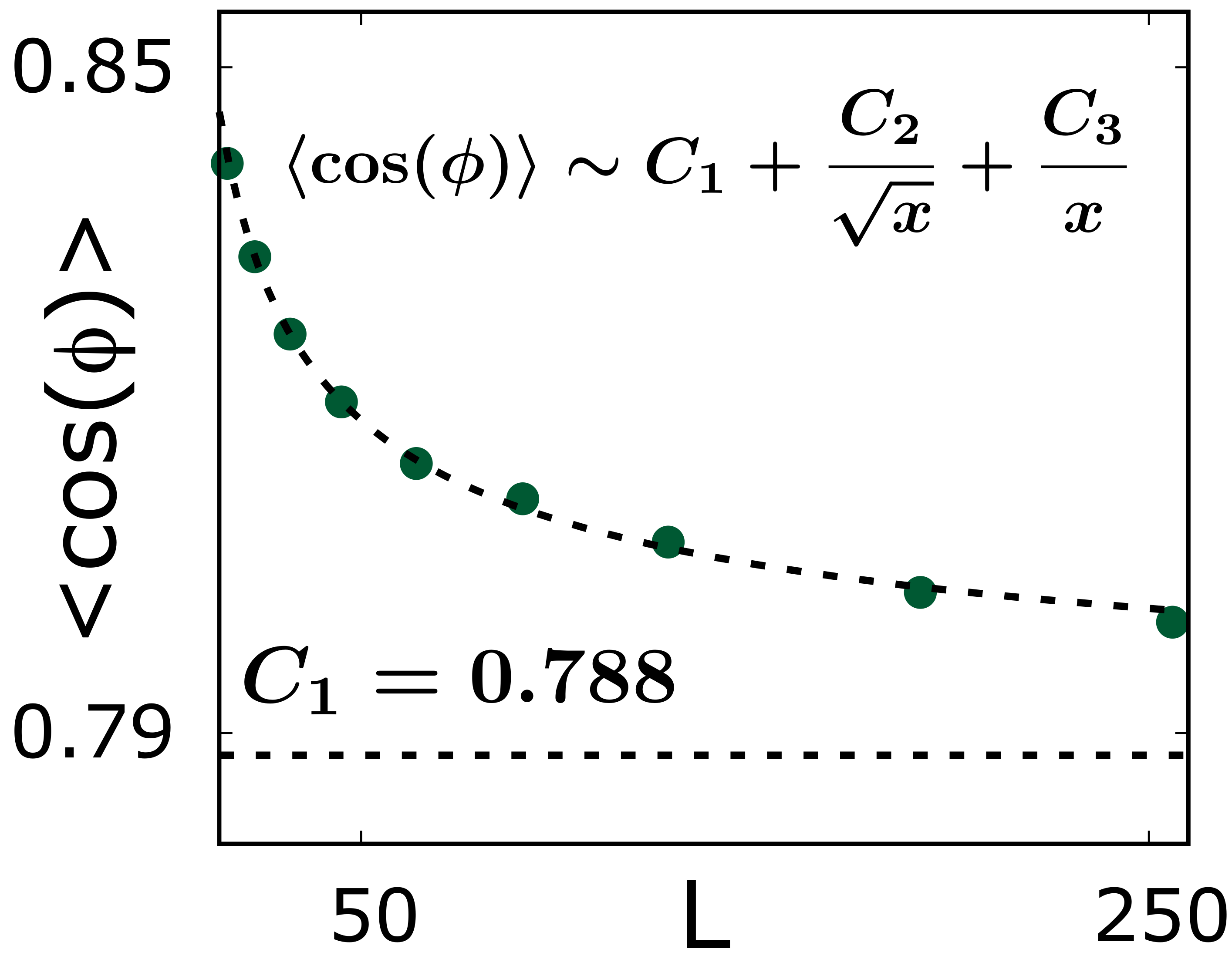}
\includegraphics[width=0.6\linewidth]{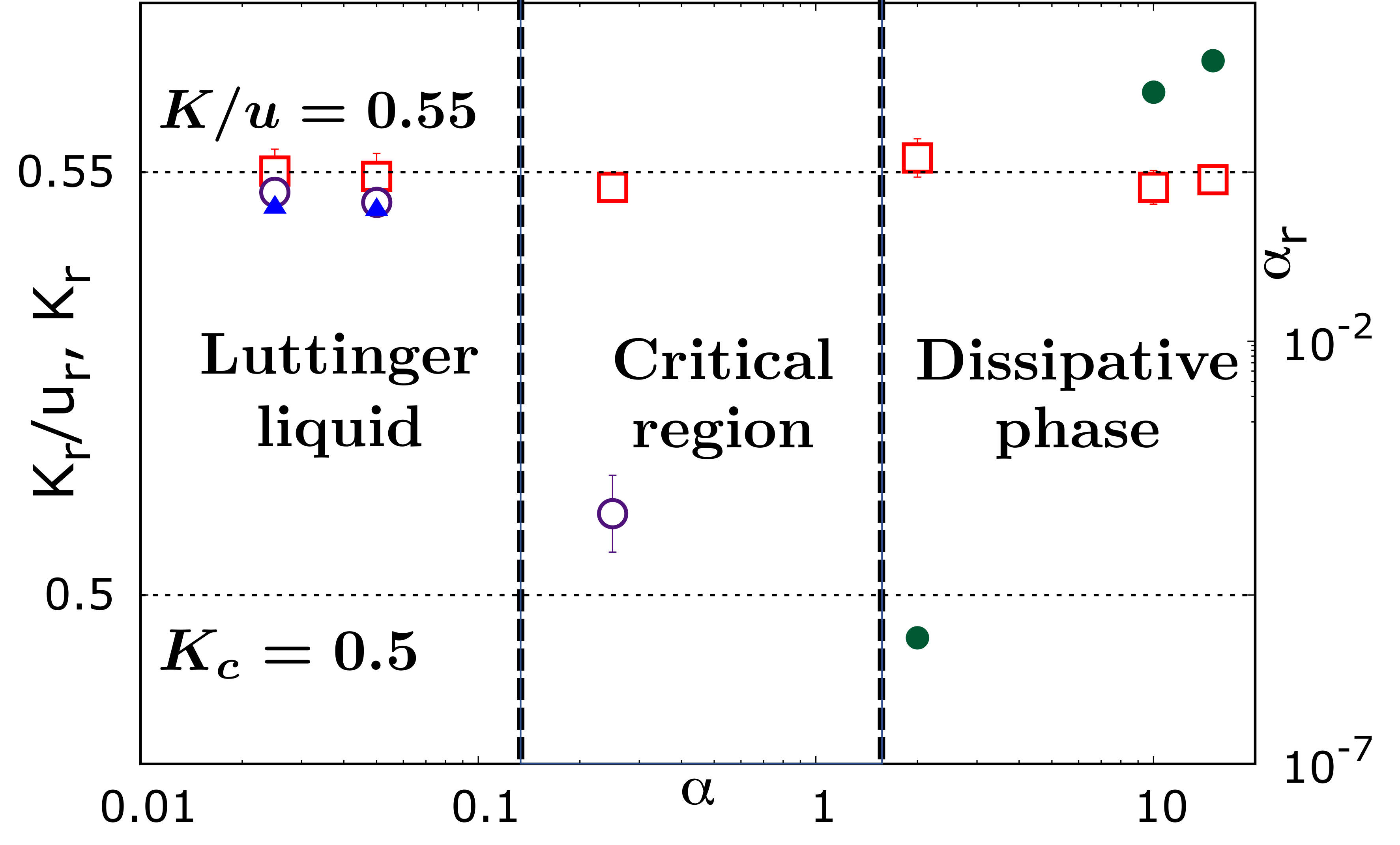}
\caption{Calculation of different quantities for $K=0.55$ that characterizes LL ($\alpha=0.05$, top row) and dissipative phase ( $\alpha=10$, bottom row).  Blue and red points correspond to $L=\beta=384$ and $L=\beta=128$, respectively.  \textit{Left} : Due to symmetry, $\pi \chi = K_r/u_r$ is equal to $K/u=0.55$ for all values of $\alpha$ and all lengthscales.  \textit{Middle} : For $\alpha=0.05$,  $\omega_n C(\omega_n)$ saturates to $K_r/2 = 0.273$ as $\omega_n \to 0$; whereas for $\alpha=10$,  $\sqrt{\omega_n} C(\omega_n)$ saturates to  $ [K_r \pi/ (8 \alpha_r u_r)]^{1/2} =0.156$.  The other fitting constants are $a_1 = 16.61$ and $a_2 = 571.4$.  \textit{Right} : For $ \alpha=0.05$, $ \langle \cos(\phi) \rangle$ decays as a power law, which allows us to extract $K_r=0.546$, consistent with the fit of $\omega_n C(\omega_n)$. For $\alpha=10$ it saturates to a constant, as predicted  by the variational ansatz (the  fit gives $c_1 = 0.788,$ $c_2 = 0.215$ and $c_3 = 0.012$).  \textit{Bottom row:} Behaviour of different renormalised parameters as a function of dissipative coupling $\alpha$.  $K_r/u_r$ (red square points) remains constant and equal to $K/u = 0.55$ for all values of $\alpha$.  $K_r$, extracted from $\langle \cos (\phi)\rangle$ analysis (purple hollow circular points) agrees with the one from the $C(\omega)$ analysis (blue solid triangular points). It approaches $K_c=0.5$ as $\alpha$ reaches the critical point. The parameter $\alpha_r$ (green solid circular points) starts to be defined in the  dissipative phase and increases rapidly with increase in $\alpha$.  This behaviour of the parameter allows to locate the different phases : For $\alpha \in [0,0.25)$, the system is in LL phase, whereas for $\alpha \in (2,15]$ the system is in dissipative phase. The phase transition takes place for $\alpha \in (0.25,2)$.  We believe $\alpha=0.25$ to be in critical region as $K_r$ extracted from $\langle \cos(\phi) \rangle$ is very close to $K_c=0.5$, $\alpha_r$ is very small and $\omega_n C(\omega_n)$ saturates for a long range of $\omega$ but then starts decreasing.  }
\label{fig:orderpar} 
\end{figure}

\end{document}